\documentclass[aps,prl,preprint,superscriptaddress]{revtex4}

\usepackage{graphicx}
\usepackage{dcolumn}
\usepackage{bm}

\begin{document}
\newcommand{\lao}{LaAlO$_3$} 
\newcommand{\sto}{SrTiO$_3$} 
\newcommand{\lsato}{La$_{1-x}$Sr$_x$Al$_{1-y}$Ti$_y$O$_3$} 


\title{Evolution of the interfacial structure of LaAlO$_3$ on SrTiO$_3$}


\author{S.A.~Pauli}
 \affiliation{Paul Scherrer Institut, CH-5232 Villigen, Switzerland}
\author{S.J.~Leake}
 \affiliation{Paul Scherrer Institut, CH-5232 Villigen, Switzerland}
\author{B.~Delley}
 \affiliation{Paul Scherrer Institut, CH-5232 Villigen, Switzerland}
\author{M.~Bj\"orck}
 \altaffiliation[Present address: ]{MAX-lab, P.O. Box 118, SE-22100
 Lund, Sweden} 
\author{C.W.~Schneider}
 \affiliation{Paul Scherrer Institut, CH-5232 Villigen, Switzerland}
\author{C.M.~Schlep\"utz}
 \altaffiliation[Present address: ]{Department of Physics,
 University of Michigan, Ann Arbor, MI 48109-1040, USA} 
\author{D.~Martoccia}
 \affiliation{Paul Scherrer Institut, CH-5232 Villigen, Switzerland}
\author{S.~Paetel}
 \affiliation{Experimental Physics VI, Center for Electronic
 Correlations and Magnetism, Institute of Physics, University of
 Augsburg, D-86135 Augsburg, Germany} 
\author{J.~Mannhart}
 \affiliation{Experimental Physics VI, Center for Electronic
 Correlations and Magnetism, Institute of Physics, University of
 Augsburg, D-86135 Augsburg, Germany} 
\author{P.R.~Willmott}
 \email[]{philip.willmott@psi.ch}
 \affiliation{Paul Scherrer Institut, CH-5232 Villigen, Switzerland}

\date{\today}

\begin{abstract}
The evolution of the atomic structure of LaAlO$_3$ grown on SrTiO$_3$ was investigated using surface x-ray diffraction in conjunction with model-independent, phase-retrieval algorithms between two and five monolayers film thickness. A depolarizing buckling is observed between cation and oxygen positions in response to the electric field of polar \lao, which decreases with increasing film thickness. We explain this in terms of competition between elastic strain energy, electrostatic energy, and electronic reconstructions. The findings are qualitatively reproduced by density-functional-theory calculations. Significant cationic intermixing across the interface extends approximately three monolayers for all film thicknesses. The interfaces of films thinner than four monolayers therefore extend to the surface, which might affect conductivity. 
\end{abstract}


\maketitle


The conducting interface between the band insulators \lao\ (LAO) and \sto\ (STO) has attracted considerable interest since its discovery in 2004\cite{Ohtomo04}. Key open questions include the origin of the conductivity associated with intrinsic doping in fully oxidized samples\cite{Thiel06,Kalabukhov07,Pentcheva09}, and why a minimum thickness of the LAO film of four monolayers (MLs) is required before the interface becomes conducting\cite{Thiel06}. 

The original explanation for the conductivity was made in terms of the buildup of a `polar catastrophe' resulting from the fact that LAO is polar, i.e., it consists of alternating positively and negatively charged layers, (LaO)$^{+}$ and (AlO$_2$)$^{-}$, while STO has charge-neutral layers\cite{Ohtomo04}. Transfer of half an electron across the interface would neutralize the buildup of electrostatic energy and thereby provide conducting electrons associated with trivalent Ti$^{3+}$. More recently, the effects of intermixing at the interface\cite{Nakagawa06,Willmott07,Chambers10} and buckling of atomic planes parallel to the interface\cite{Pentcheva09} have been proposed as contributory factors. 

A common feature of many perovskites is that structural changes as small as $0.1$~\AA\ or less can induce fundamental changes in their physical properties\cite{Dagotto05}. A knowledge of the structural subtleties with sufficient accuracy can therefore be invaluable in elucidating the underlying physics. Surface x-ray diffraction (SXRD) can offer this level of structural resolution\cite{Feidenhansl89}. In this Letter, we describe the evolution of the interfacial structure of LAO on STO as a function of LAO film thickness, determined by SXRD in conjunction with phase-retrieval algorithms, and show how competing energetic factors lead to the formation of conductivity at the interface. 


Films of $2$, $3$, $4$, and $5$-ML thickness were prepared by pulsed laser deposition using standard growth conditions\cite{Schneider06}. The samples were subsequently checked by atomic-force microscopy (AFM) for atomic flatness. SXRD measurements were performed at room temperature at the Materials Science beamline, Swiss Light Source, Paul Scherrer Institut, using $16$~keV ($0.775$~\AA) photons. For each film thickness $15$~inequivalent crystal truncation rods (CTRs) were recorded up to a scattering vector of 11.3~\AA$^{-1}$ using the PILATUS~100k pixel detector\cite{Schlepuetz05}. Additional symmetry-equivalent CTRs were also recorded to obtain the systematic errors of approximately $5$~\%. The data were analyzed using the DCAF phase-retrieval algorithm\cite{Bjorck08} to obtain average electron-density maps\cite{footnote1}, which were used as starting models for further structural refinement with the grid-search $\chi^2$-minimization program \textsl{fit}\cite{Bunk99}. In total $N+5$ unit cells were taken into account for the refinement, where $N$ is the number of LAO MLs. Each atom was fit for its position and isotropic Debye-Waller factor. Additional fit parameters included partial occupations of the A- and B-sites (La/Sr and Ti/Al, respectively) as well as the occupations of the top two unit cells. The final models exhibited \textsl{R}-factors of $5.5~\%$, $7.5~\%$, $7.0~\%$, and $6.6~\%$ for the $2$, $3$, $4$, and $5$-ML data sets, respectively. 

\begin{figure}
\includegraphics[scale=0.345]{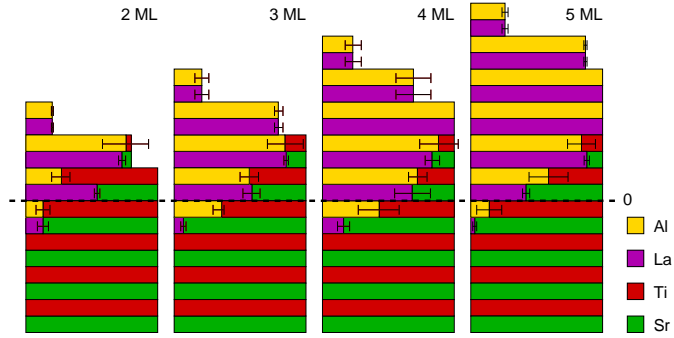}
\caption{\label{fig:occ}(color) Refined cation occupation for all four film
  thicknesses depicted as blocks. Intermixing of more than
  approximately $5$~\% extends across $3$~ML at the interface for
  all films, which also exhibit the same apparent partial occupation
  of the top two MLs of approximately $80$ and $20$~\%. The horizontal
  line at 0 marks the nominal interface.} 
\end{figure}

Fig.~\ref{fig:occ} shows the refined occupations. There is a consistent coverage of approximately $80~\%$ for the nominally top layer, plus another $20~\%$ coverage on top of that, despite the fact that no isolated islands or gaps in the coverage could be established in AFM images of both the STO substrates before growth and the films after growth. This can be simply explained as being due to a small lateral gradient of the film thickness across the substrate\cite{Willmott00}. Another more intriguing possibility runs as follows. The fact that even films of considerably greater thicknesses exhibit atomically smooth terraces and straight terrace edges would seem to imply that growth is partially dictated by a step-flow mode. This means that the terrace edges can drift laterally. 
Hence there can be parts of a film of nominally $N$~ML that have thicknesses of $N+1$ or $N-1$~ML, even if the surface of the film shows no islands or wells. As x-rays penetrate the entire film thickness, the SXRD data reflect this variation in thickness. 

According to the SXRD results, cationic intermixing greater than approximately $5$~\% extends across three monolayers at the interface for all four measured thicknesses. Smaller degrees of intermixing may extend even further into the substrate and the film, as reported by Qiao \textsl{et al.}\cite{Chambers10}, although this is below our experimental sensitivity. The refined structures of the two- and three-ML samples contain some Sr and Ti atoms at the surface. 


\begin{figure*}
\includegraphics[scale=0.515]{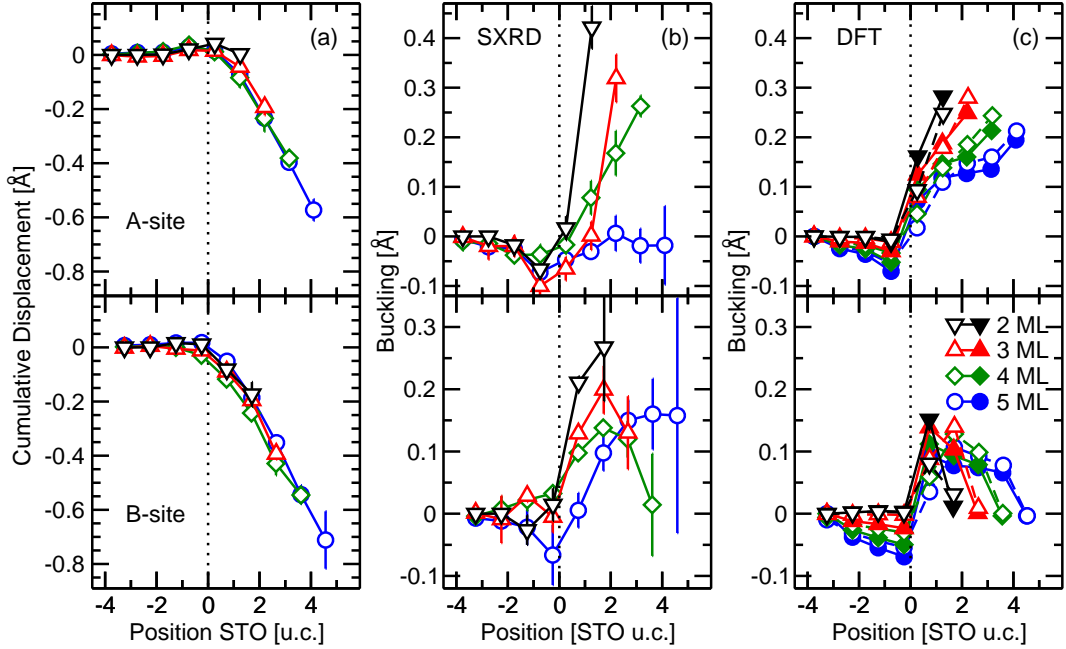}
\caption{\label{fig:structure}(color online) (a) Cumulative displacement out of plane of the atomic positions relative to a reference grid defined by bulk STO. For reasons of clarity, only the average of the A-site (upper panels) and B-site(lower panels) atomic layer positions are shown. (b) and (c) show the buckling of the A-site and B-site atomic planes from the refined structure and the DFT calculations, respectively, shown on the same scale. Positive values indicate movements of the cation relative to the oxygen ions towards the surface. In (c) the filled and open circles mark the abrupt and intermixed DFT models, respectively. The dotted lines represent the nominal interface.}
\end{figure*}

Pertinent features of the structures are summarized in Fig.~\ref{fig:structure}. The less reliable values associated with the top $20$~\% coverage are not included in order to display the results on the same scale. The films are perfectly strained in plane. The out-of-plane lattice constant of the LAO layers above the intermixed interface is $3.73 \pm 0.01$~\AA, consistent with a Poisson ratio of $0.24$ for LAO\cite{Luo08}. The average atomic layer positions for the A- and B-sites are shown in Fig.~\ref{fig:structure}(a). For the A-site, and to a smaller extent also for the B-site, we see an increase in the c-lattice constant of STO as it approaches the nominal interface. This is attributed to substitutional incorporation of La cations, and/or the presence of Ti$^{3+}$ atoms\cite{Willmott07}.

Recently, buckling of the atomic layers in LAO was predicted by density functional theory (DFT)\cite{Pentcheva09}. Dipole moments are induced in opposition to the electric field of the polar film layers. Little change in the amplitude of the buckling as a function of film thickness was observed. Our experimentally determined structures exhibit a qualitatively similar positive buckling in the films [see Fig.~\ref{fig:structure}(b)], whereby positive buckling is defined by the cation moving towards the surface relative to the oxygen atom. However, buckling in the A-site layers is more pronounced for the $2$-ML film than was predicted by DFT, and also drops off significantly with film thickness. Buckling at the B-site also decreases with film thickness, though less pronouncedly. Interestingly, the near-interface region of the STO exhibits a negative buckling, as predicted in \cite{Schwingenschloegl09b}. In contrast to the film buckling, this {\em increases} with increasing layer thickness. 

To further investigate the behavior of the buckling, we performed DFT calculations for all four thicknesses using the local density approximation\cite{Delley00,Delley02}. The substrate consisted of a lower $3\times$(TiO$_2$)/$2 \times$(SrO) layers fixed at the calculated DFT bulk positions, plus $3\times$(TiO$_2$)/$3\times$(SrO) layers which were allowed to relax. Two models were investigated for which the results are shown in Fig.~\ref{fig:structure}(c). The first model assumed an abrupt interface (i.e., one with no intermixing). For both the A- and B-sites, there is a consistent reduction in the positive buckling with increasing film thickness, in qualitative agreement with our experimental findings, and also negative buckling in the substrate close to the surface, which increases with the layer thickness. The most notable difference is the collapse of the buckling for the A-site found experimentally for the $5$-ML film, which however, is still evident in the DFT results. 

Since our experimental findings showed that the interface is not sharp, the influence of intermixing on the buckling was also studied. We therefore investigated a second model with DFT in which the bottom unit cell of the film contained 50~$~\%$ LAO and 50~$~\%$ STO occupation. This causes a reduction of the buckling magnitude close to the intermixed layer, while above the nominal interface, buckling is marginally greater than that for the abrupt model -- both these changes are in better agreement with our experimental findings. 

In the simplest picture of the polar model of LAO on STO, one can describe the band scheme of the LAO film in terms of a simple plate capacitor, with a positively charged layer at the LAO/STO interface and a negative layer at the surface. The electric field in between the two ``plates'' is constant, and the potential therefore increases linearly with film thickness. In the framework of the polar-catastrophe model, electrons from the film surface move across the film to the interface once the film thickness is large enough that the valence band crosses the Fermi level. 

\begin{figure}
\includegraphics[scale=0.31]{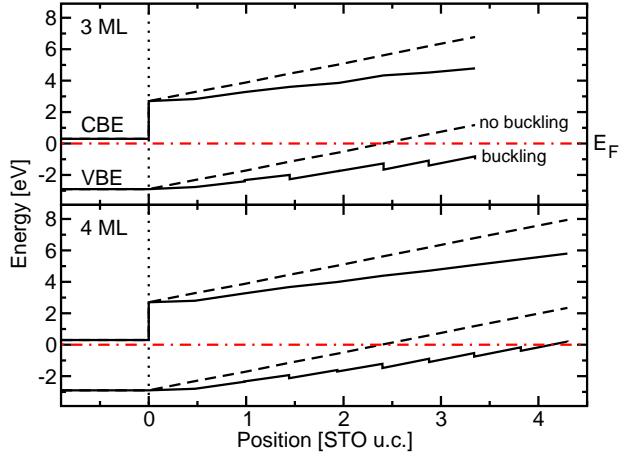}
\caption{\label{fig:bandstructure} (color online) The influence of buckling on the valence band edge (VBE) and conduction band edge (CBE) relative to the Fermi level ($E_F$) for the 3-ML- and 4-ML-LAO films. Without buckling, the electric field is constant across the film. Buckling results in a zigzag motif, shown for the VBE. Negative buckling in the STO and the partial occupation at the surface were taken into account, but are not shown for the sake of clarity. Since from our ``ionic'' model, we cannot determine band bending in STO, this was neglected.}
\end{figure}

We have calculated the influence of buckling on this simple description of the band scheme. Figure~\ref{fig:bandstructure} shows the results for $3$ and $4$~ML. Buckling is induced as a depolarizing effect to reduce the potential within the film and thereby increases the minimum thickness at which the electronic reconstruction occurs, by lowering the average gradient of the potential within the film. Once the valence band moves above the Fermi level, however, electron injection across the interface occurs, causing the ``capacitor'' to discharge. The potential collapses and obviates the need for a depolarizing buckling. Using our experimentally determined atomic structures, and assuming formal charges for the cations and oxygen ions, we see that this occurs at $4$~ML -- the valence band moves across the Fermi level and the positive buckling in the film, particularly for the A-site, collapses and is essentially zero for the $5$-ML sample. 

We now address the negative buckling of the STO just below the nominal interface\cite{Schwingenschloegl09b}. The electrons injected across the interface are confined to near the interface in the STO by band bending in this region. The gradient in the band bending region results in a potential in the opposite direction to that in the film. This causes negative buckling of the STO layers once the 2-dimensional conducting layer is formed, as also seen experimentally. 

Buckling costs elastic energy, as given by 
\begin{eqnarray} 
E = Y a \left(B_A^{\phantom{A}2} + B_B^{\phantom{B}2}\right),
\end{eqnarray}
per ML, whereby $Y$ is the Young's modulus, $a$ is the STO lattice constant, and $B_A$ and $B_B$ are the A-site and B-site out-of-plane buckling amplitudes, respectively. Based on calculations of deviations from a ``start model'' of the known stoichiometry and a Young's modulus of $Y = 306$~GPa for LAO\cite{Luo08}, the energy cost per ML and a buckling of $0.2$~\AA\ is $0.59$~eV. On the other hand, the electrostatic energy gain per unit cell is given by 
\begin{eqnarray} 
e\Delta V = e\frac{q_A B_A+q_B B_B}{\epsilon \epsilon_0 a^2},
\end{eqnarray}
with $q_A$ and $q_B$ equal to the ionic charges, $B$ the buckling, and the relative permittivity $\epsilon=24$\cite{footnote2}. With the simplification $q=3e$ and $B_A = B_B$, the ratio between these two competing energies is 
\begin{eqnarray} 
\frac{E}{e\Delta V} & = & \frac{\epsilon \epsilon_0 Y a^3}{3e^2}\,B \\ 
                   & = & B/0.2, 
\end{eqnarray}
whereby $B$ is in \AA. In other words, buckling much in excess of $0.2$~\AA\ becomes energetically unfavorable. Both our experimental and DFT results comply well with this energetic constraint. 

The DFT results show only modest differences between the abrupt and $1$-ML intermixed models with regards to the buckling and partial density of states (not shown). Potentially more significant differences associated with more extensive intermixing would be very difficult to investigate with DFT because of the unrealistic computational effort. Hence we cannot completely exclude intermixing from playing a role in the formation of the conducting layer, although {\em per se} it cannot easily explain why n-type interfaces are conducting, but those of p-type are insulating. 

On the other hand, conductivity has been observed in LAO layers thinner than $4$~ML if they are capped with a sufficient thickness of STO\cite{Cancellieri10,Pentcheva10}. Our structural analysis demonstrates that for thicknesses of $3$~ML and less, the uppermost layer is significantly intermixed. Only for $4$~ML and above is the interface electrically isolated from the surface with one or more complete MLs of LAO containing an intermixed fraction of less than $5$~\%, the approximate limit to the sensitivity of SXRD. It can therefore be speculated that within the framework of the intermixing model, surface effects could influence the conductivity of LAO layers thinner than $4$~ML, which might also explain why capping ultrathin LAO layers with STO preserves the conductivity. 

In conclusion, using SXRD, phase-retrieval methods and subsequent fitting, we have solved the atomic structures of \lao\ grown heteroepitaxially on \sto\ for $2$, $3$, $4$, and $5$~ML with a resolution better than $0.1$~\AA, even for the oxygen-atom positions. Buckling of the cation-oxygen planes in the LAO films is strongest for the thinnest $2$-ML LAO layer and decreases with increasing film thickness. This behavior has been explained as a response to the internal electric field generated by the polar nature of LAO. DFT calculations qualitatively reproduce these results. More modest buckling in the opposite direction is also observed in the uppermost STO layers, which increases with film thickness in response to the injection of electrons across the interface. The refined structures consistently exhibit a nonabrupt interface with cationic intermixing extending over three monolayers. The interfaces of the $2$- and $3$-ML samples hence extend to the surface, which may influence the electronic properties. 

\begin{acknowledgments}
Support of this work by the Schweizerischer Nationalfonds zur F\"orderung der wissenschaftlichen Forschung and the staff of the Swiss Light Source is gratefully acknowledged. JM and S.~Paetel also gratefully acknowledge financial support from the DFG~TRR~80 and EU~Oxides programs. This work was partly performed at the Swiss Light Source, Paul Scherrer Institut.
\end{acknowledgments}

\bibliography{bibliography}

\begin{thebibliography}{22}
\bibitem[{foo({\natexlab{a}})}]{footnote1}
\bibinfo{note}{For each thickness $100$ different initial guesses seeded with
  random phases or random electron-density were run. After convergence,
  Gaussians were fit to the peaks in the electron-density maps. Each atomic
  position and occupation from the 100 runs were averaged, which also yielded
  the error margins.}
\bibitem[{foo({\natexlab{b}})}]{footnote2}
\bibinfo{note}{This bulk value for the relative permittivity of LAO may be an
  overestimate, because of intermixing of LAO with STO for which there is no
  reliable data.}
\end{thebibliography}
\end{document}